\documentclass[11pt, onecolumn, twoside, a4paper]{article}
\usepackage{iiii-j2}
\usepackage{amsmath}
\usepackage{amssymb}
\usepackage{amsthm}
\usepackage{mathptmx}

\usepackage[rm,up,sc,topmarks,calcwidth,pagestyles]{titlesec}
\titleformat{\section}{\Large\bfseries\rmfamily}{\thesection}{1em}{}
\titleformat{\subsection}{\large\bfseries\rmfamily}{\thesubsection}{1em}{}
\titleformat{\subsubsection}{\large\it\rmfamily}{\thesubsubsection}{1em}{}

\usepackage{bm}
\usepackage{amsthm}
\theoremstyle{Definition}

\newtheorem{thm}{Theorem}

\usepackage{url}

\usepackage{cite}
\usepackage{algorithmic}
\usepackage[dvipdfmx]{graphicx}
\usepackage{textcomp}

\usepackage{color}

\authorshead{H.~Hirose}
\titlehead{Matrix Decomposition Perspective for Accuracy Assessment of Item Response Theory\\}
\vol{ }
\no{ }
\year{ }

\title{Matrix Decomposition Perspective  \\ 
for Accuracy Assessment
of Item Response Theory
\\}
\author{Hideo Hirose
 \thanks{Kurume University, Fukuoka, Japan}
}
\date{}
\begin{document}
\maketitle
\thispagestyle{empty}

\section*{Abstract}

The item response theory obtains the estimates and their confidence intervals for parameters of abilities of examinees and difficulties of problems by using the observed item response matrix consisting of 0/1 value elements. Many papers discuss the performance of the estimates. However, this paper does not. Using the maximum likelihood estimates, we can reconstruct the estimated item response matrix. Then we can assess the accuracy of this reconstructed matrix to the observed response matrix from the matrix decomposition perspective. That is, this paper focuses on the performance of the reconstructed response matrix. 

To compare the performance of the item response theory with others, we provided the two kinds of low rank response matrix by approximating the observed response matrix; one is the matrix via the singular value decomposition method when the response matrix is a complete matrix, and the other is the matrix via the matrix decomposition method when the response matrix is an incomplete matrix. 

We have, firstly, found that the performance of the singular value decomposition method and the matrix decomposition method is almost the same when the response matrix is a complete matrix. Here, the performance is measured by the closeness between the two matrices using the root mean squared errors and the accuracy. Secondary, we have seen that the closeness of the reconstructed matrix obtained from the item response theory to the observed matrix is located between the two approximated low rank response matrices obtained from the matrix decomposition method of $k=1$ and $k=2$ to the observed matrix, where $k$ indicates the first $k$ columns use in the decomposed matrices. 

This is amazing. The reconstructed response matrix using the item response theory, whether continuous or discrete, would not exceed the approximated low rank response matrix of $k=2$ using the singular value decomposition method or the matrix decomposition method, even if the probability function is a nonlinear function of parameters.
This perspective has been supported by many example cases. 
\\[2mm]
{\it Keywords: }singular value decomposition, matrix decomposition, item response theory, maximum likelihood estimation, root mean squared error, accuracy, matrix rank.

\section{Introduction}

The item response theory\cite{Ayala, Baker, Hambleton91, HandbookIRTtheory}, IRT, is regarded as the standard method to evaluate abilities of examinees and difficulties of problems simultaneously in testing. Thus, a variety of official tests such as TOFLE adopt the IRT in assessing the performance. To a provided item response matrix, we can obtain the estimates and their confidence intervals for parameters by using the maximum likelihood estimation method. Usually, the item response matrix consists of 0/1 valued elements and all the elements are fully occupied; that is, the matrix is a complete matrix. 

The maximum likelihood estimators are known to be consistent and asymptotically efficient under certain conditions\cite{AdvancedTheoryofStatistics}; that is, no consistent estimators have lower asymptotic mean squared errors other than the maximum likelihood estimators. 
However, this holds 
under some restricted mathematical model and its parameter space. 
In the IRT, the number of unknown parameters are $m+2n$, where $m$ denotes the numbers of examinees and $n$ the number of problem items when we use a logistic probability distribution with an unknown location parameter and a scale parameter.

By using the estimated parameters, we can reconstruct the estimated item response matrix whose element values are located in $[0,1]$. Then, we can evaluate how close the estimated response matrix is to the observed response matrix by using the root mean squared errors, RMSE. In addition, if we transform the estimated element value to 0/1 discrete value, we can also evaluate how close the transformed response matrix is to the observed response matrix by using the accuracy value, where the accuracy is the ratio of the number of correctly estimated elements to the number of total elements. However, we cannot know how well the estimated response matrix is approximated to the observed response matrix only by looking at the RMSE value and accuracy value obtained by the IRT alone. We need such values via some different mathematical model for comparison.

In this paper, we have compared the RMSE and accuracy values obtained by using the IRT to the observed item response matrix with those obtained by using the singular value decomposition, SVD, to the target response matrix. Here, the target response matrix means the approximated low rank matrix obtained from the SVD. That is, we can look at the IRT from the perspective of the SVD. As a result, we have known that the model complexity of the IRT is explained in terms of the target matrix rank obtained from the SVD. Computing methods to obtain the SVD have been proposed such as \cite{AllenGrosenickTaylor, MahoneyaDrineas, Berry}. However, this paper aims at obtaining another look at the IRT from the SVD perspective.

A typical examination case showed us an amazing result. When the matrix size is $(m, n)=(216, 31)$, the RMSE/accuracy of the reconstructed response matrix obtained from the IRT to the observed matrix is larger/smaller than that obtained from the approximated matrix by using the SVD with rank of 2 to the observed matrix, but smaller/larger than that obtained from the approximated matrix by using the SVD with rank of 1, although the rank of the observed response matrix is 31.

Sometimes, we encounter cases such that all the elements in the observed response matrix are not occupied; that is, we tackle the incomplete matrix with missing data. The IRT still can be applied in such situations by modifying the estimation methodology\cite{EeL2011, hirose2012item, IPSJ2014}. However, the SVD cannot. In such cases, the matrix decomposition method, or other methods, e.g., the imputation\cite{Troyanskaya} method with EM algorithm\cite{Dempster}, can be used instead.

\section{Item Response Matrix}

\subsection{Observed item response matrix}

\begin{figure}[htb]
\begin{center}
\includegraphics[scale=1.0]{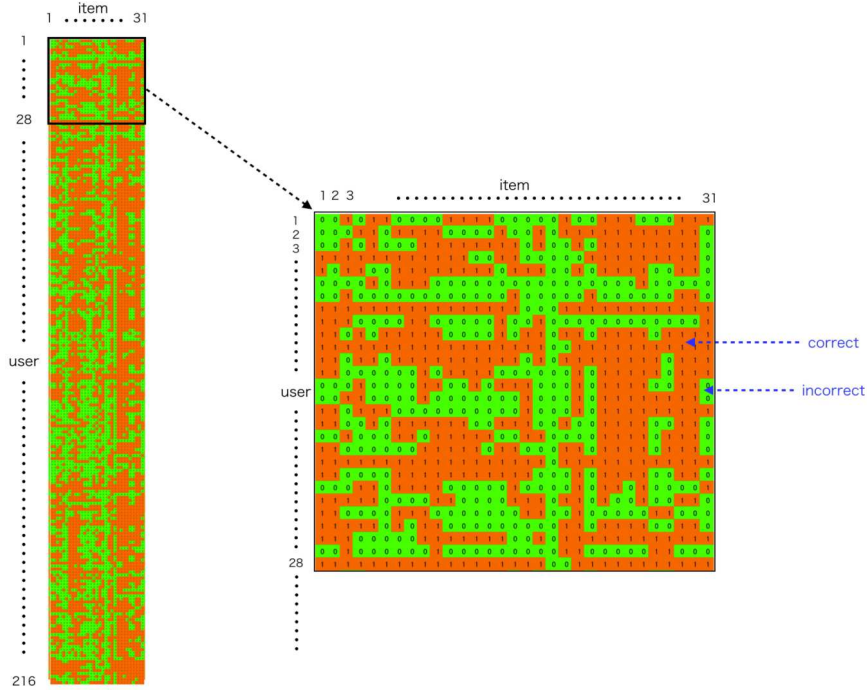}
\end{center}
\caption{Observed item response matrix.} 
\label{fig:observedmatrix}
\end{figure}

As a typical case, we use an observed item response matrix obtained from a mid-term mathematics examination test result, where the number of examinees is 216 and the number of questions is 31. 
Figure \ref{fig:observedmatrix} on the left shows the observed item response matrix; on the right, a partially extracted part from the matrix is seen for ease of viewing. 
The matrix consists of two-valued elements: 0 for incorrectly answered, and 1 for correctly answered.
The matrix is a complete matrix.
We denote the observed item response matrix as $A=(a_{ij})$.

\subsection{Estimated item response matrix}

In this paper, we intend to find where the IRT performance is in terms of accurate estimation to the observed item response matrix. Here, we define two types of matrices to represent the target response matrix: one is to use continuous values to each element in the matrix, and the other 0/1 discrete values. We call the former the continuous response matrix, and the latter the discrete response matrix. 
We denote the estimated continuous response matrix as $S=(s_{ij})$, and the estimated discrete response matrix as $T=(t_{ij})$.

Using the estimated continuous response matrix, we can evaluate how close the estimated response matrix is to the observed response matrix by using the RMSE. Similarly, using the estimated discrete response matrix, we can evaluate how close the estimated response matrix is to the observed response matrix by using the accuracy value. These values will be defined later.

\section{Item Response Theory}

\subsection{Parameter estimation}

We assume probability $P_{ij}$ that examinee $i$ answered question $j$ correctly is denoted as 
\begin{eqnarray} \nonumber
P_{ij}(\theta_i;a_j,b_j)&=&{1 \over 1+\exp\{-1.7a_j(\theta_i-b_j)\} },\\
&=&1-Q_{ij}(\theta_i;a_j,b_j),
\end{eqnarray}
where $\theta_i$ expresses the ability for examinee $i$, and $a_j, b_j$ are constants in the logistic function for question $j$ called the discrimination parameter and the difficulty parameter, respectively.
$Q_{ij}$ is the probability that examinee $i$ answered question $j$ incorrectly.
Then, we can obtain the maximum likelihood estimates $\hat{\theta}_i$ and $\hat{a}_j, \hat{b}_j$ for parameters $\theta_i$ and $a_j, b_j$ by maximizing the likelihood function,
\begin{eqnarray} 
L=\prod_{i=1}^m \prod_{j=1}^n \left(P_{ij}^{\delta_{ij}} 
\times Q_{ij}^{1-\delta_{ij}} \right),
\end{eqnarray}
where $m$ and $n$ are the number of examinees and the number of questions, respectively, and $\delta_{ij}$ is the indicator function such that $\delta_{ij}=1$ if examinee $i$ solved item $j$ successfully and $\delta_{ij}=0$ otherwise.
We also call examinees as users, and questions as items, in terms of recommender systems.

It is not easy to obtain the estimates for item parameters and users' abilities together simultaneously because $\theta_i$ and $b_j$ could be drifting preserving the same value of $\theta_i-b_j$.
Therefore, item parameters $a_j$ and $b_j$ are first estimated by using the marginal likelihood function by removing examinees' abilities such as 
\begin{eqnarray}  
L(\textrm{\boldmath $\delta$}|a,b)=\prod_{i=1}^m \large[ \int_{-\infty}^{\infty} g(\theta) \prod_{j=1}^n L(\delta_{ij}|a_j,b_j)d\theta \large],
\end{eqnarray}
where $g(\theta)$ denotes the prior to the ability distribution common to all the examinees and $\textrm{\boldmath $\delta$}$ denotes all the patterns of $\delta_{ij}$ taking 0/1 value. 
We often apply $g(\theta)$ to a standard normal distribution. 
The EM algorithm \cite{Dempster} is usually used in such a case \cite{Baker}.  
After $a_j$ and $b_j$ are determined, examinees' abilities $\theta_i$ can be obtained by maximizing the corresponding likelihood function. 
To circumvent ill conditions such that all the questions are correctly answered or incorrectly answered, a Bayes technique is applied \cite{Baker}. 

Usually, the function $\delta_{ij}$ takes values such that $\delta_{ij}=1$ for success and $\delta_{ij}=0$ for failure in answering a question. 
However, we can extend $\delta_{ij}$ value from 0/1 discrete value to a continuous value in $[0, 1]$ corresponding to the answering level, and a null value in element $(i,j)$ can be dealt with when a value of element $(i,j)$ is vacant \cite{EeL2011, hirose2012item, IPSJ2014}. This corresponds the case that examinee $i$ did not tackle problem $j$. 

\subsection{Construction of the estimated response matrix}

Once the estimates for the parameters are obtained, we can reconstruct the continuous response matrix by using the estimated $\hat{P}_{ij}(\hat{\theta}_i;\hat{a}_j,\hat{b}_j)$ to each element directly; that is $s_{ij}=\hat{P}_{ij}(\hat{\theta}_i;\hat{a}_j,\hat{b}_j)$. If we transform the estimated $\hat{P}_{ij}(\hat{\theta}_i;\hat{a}_j,\hat{b}_j)$ to 0/1 discrete value such that $t_{ij}=0, \ (\hat{P} < 0.5), \ t_{ij}=1, \ (\hat{P} \ge 0.5)$, we can construct the discrete response matrix.

\section{Singular Value Decomposition}

\subsection{Procedure to make the target response matrix}

Assuming that $A$ is a $m \times n$ matrix. Then, $A^\mathsf{T} A$ becomes a $n \times n$ symmetric matrix, and $A A^\mathsf{T}$ becomes a $m \times m$ symmetric matrix, where $A^\mathsf{T}$ denotes the transpose of $A$. Eigen values and eigen vectors to these two matrices $A^\mathsf{T} A$ and $A A^\mathsf{T}$ are the same if they exist.
We denote the eigen values and eigen vectors to matrix $A^\mathsf{T} A$ as $\{\xi_1, \xi_2, \cdots , \xi_{n}\}$ and $\{ \bm{v}_1, \bm{v}_2, \cdots, \bm{v}_n \}$. 
That is, 
\begin{eqnarray} 
A^\mathsf{T} A \bm{v}_i = \xi_i \bm{v}_i.
\end{eqnarray}

Eigen values can be reordered such that $\xi_1 \ge  \xi_2 \ge  \cdots \ge  \xi_r > 0, \xi_{r+1}= \cdots =\xi_{n}=0$, where $r$ is the rank of $A^\mathsf{T} A$. Since $A^\mathsf{T} A$ is symmetric, eigen vectors can be made as orthonormal system. That is, $\bm{v}_i \cdot \bm{v}_j = \delta_{ij}$, where $\delta$ is the indicator function. 
We make vector $\bm{u}_i$ by $\bm{u}_i = A \bm{v}_i / {\sigma_i}, \ (i \ge r)$, 
where ${\sigma_i} = \sqrt{\xi_i}$. 
In addition, if we produce matrices $U=(\bm{u}_i)$ and $V=(\bm{v}_j)$, then $A$ can be expressed as $A = U \Sigma V^\mathsf{T}$, or equivalently, 
$A=\sum_{l=1}^r \sigma_l \bm{u}_l \bm{v}^\mathsf{T}_l$. 
Here, $\Sigma$ is a diagonal matrix using ${\sigma_i}$. 
This is the singular value decomposition\cite{Golub, StrangMultiplyingFactoringMatrices, StrangLinearAlgebra}. 

We define $A_k$ such that 
$
  A_k  = \sum_{l=1}^k \sigma_l \bm{u}_l  \bm{v}_l ^\mathsf{T},
$
using the first $k$ columns in the matrices of $U$ and $V$.
This is the procedure to make the target response matrix.

\subsection{Matrix approximation by using the SVD}

It is interesting to show 
the following theorem\cite{EckartYoung}.
\begin{thm}[Eckart-Young]$\\$
1) $ rank(A_k ) = k$$\\$
2) For any $m \times n$ matrix $B, \ (rank(B) \le k)$, 
\end{thm}
$
\ \ \ \  \displaystyle 
  || A - A_k || = \min_{B, rank(B) \le k} || A - B || =  (\sum_{l=k+1}^{n} \sigma^2_{l})^{1/2}
$, 

\noindent
where $|| \cdot ||$ means the Frobenius matrix norm. 

The theorem claims that $A_k$ is best approximated to $A$ among all the matrices with rank of less than $k+1$ in the sense of matrix norm. 

\subsection{Construction of the target response matrix}

We use $S$ instead of $A_k$ to all $k$ here. 
We can construct the continuous response matrix with rank of $k$ by using the SVD to the observed response matrix; that is $s_{ij}=\sum_{l=1}^k \sigma_l \bm{u}_l  \bm{v}_l ^\mathsf{T}$. If we transform $s_{ij}$ to 0/1 discrete value such that $t_{ij}=0, \ (s_{ij} < 0.5), \ t_{ij}=1, \ (s_{ij} \ge 0.5)$, we can construct the discrete response matrix.

\section{Matrix Decomposition}

\subsection{Procedure to make the target response matrix}

The SVD is a promising method to find approximate matrices close to a certain matrix in a sense of the matrix norm. However, it requires that all the elements are occupied. 
Sometimes, we encounter cases such that all the elements in the observed response matrix are not occupied. In such cases, the matrix decomposition method\cite{KorenBellVolinsky}, or other methods, e.g., the imputation method with EM algorithm, can be used instead.

The matrix decomposition is similar to the SVD. We can construct a target matrix $R \in \mathbb{R}^{m \times n}$ with two matrices $U \in \mathbb{R}^{m \times k}$ and $V \in \mathbb{R}^{n \times k}$, but there are no singular value matrix or eigen value matrix in the decomposition form. Matrix decomposition is described as
\begin{eqnarray} 
R = U V^\mathsf{T}.
\end{eqnarray}
Using the non-null element values in $A$, we find $U$ and $V$ so that
\begin{eqnarray}
 E=\sum_{i=1}^{m}{\sum_{j=1}^{n}{I(i,j)(a_{ij}-r_{ij})^2}}
\end{eqnarray}
becomes small, where $r_{ij}=\sum_{l=1}^k u_{il} v_{jl}$, and $I(i,j)$ is the indicator function such that $I(i,j)=1, \ \text{if $a_{ij}$ is non-null}, \ I(i,j)=0, \ \text{if $a_{ij}$ is null}$. For stable computation, we use another function such that 
\begin{eqnarray}
\nonumber
W=\sum_{i=1}^{m}{\sum_{j=1}^{n}{I(i,j)(a_{ij}-r_{ij})^2}} 
+ {k_u} \sum_{i=1}^{m} \sum_{l=1}^{k} u_{il} ^2 
   +   {k_v} \sum_{j=1}^{n}  \sum_{l=1}^{k} v_{jl} ^2, 
\end{eqnarray}
where, ${k_u}$ and ${k_v}$ are regularization coefficients to prevent overfitting.
To find the optimum values, we use the descent method\cite{Fletcher, Polak}.
From appropriately set initial values of $u^{(0)}_{il}$ and $v^{(0)}_{jl}$, we proceed the following iteration until $|u^{(t+1)}_{il} - u^{(t)}_{il}|$ and $|v^{(t+1)}_{jl} - v^{(t)}_{jl}|$ are both small enough.
\begin{eqnarray}
\nonumber
u^{(t+1)}_{il} &\leftarrow& u^{(t)}_{il} - \mu \frac{\partial W}{\partial u_{il}}|^{(t)} \\ 
v^{(t+1)}_{jl} &\leftarrow& v^{(t)}_{jl} - \mu \frac{\partial W}{\partial v_{jl}}|^{(t)},
\end{eqnarray}
where, $\mu$ is a learning coefficient.

\begin{figure}[htb]
\begin{center}
\includegraphics[scale=1.0]{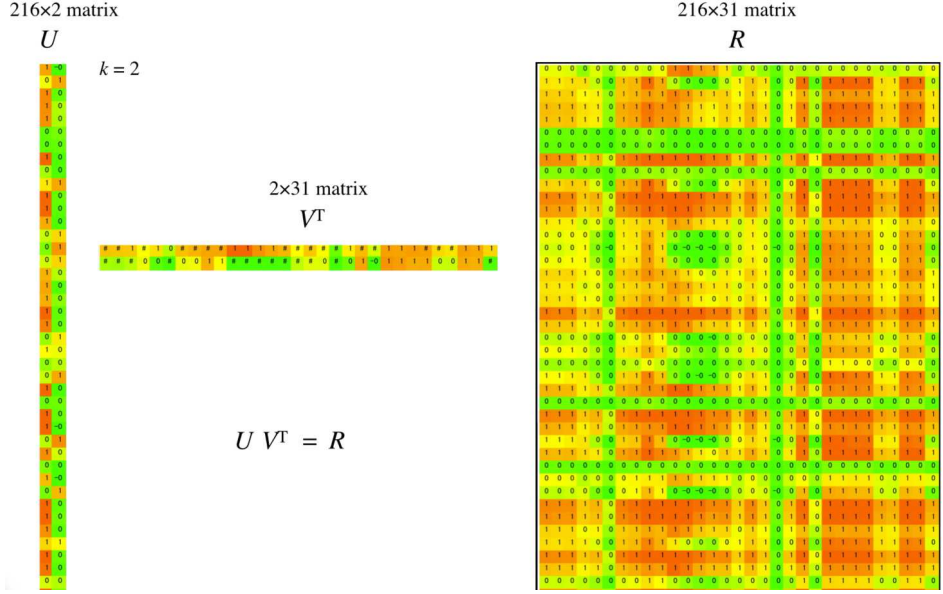}
\end{center}
\caption{Matrix decomposition of the observed response matrix when $k=2$.} 
\label{fig:UVR2}
\end{figure}

Figure \ref{fig:UVR2} shows a procedure that matrix multiplication of $U$ and $V^\mathsf{T}$ makes $R$ when $k=2$ in the matrix decomposition method. For clarity, the figure shows parts of the matrices.
It is interesting that even only two vectors use in $U$ and $V$ can configure a rather clear matrix approximated to the observed response matrix shown in Figure \ref{fig:observedmatrix}.

\subsection{Construction of the target response matrix}

We can choose $k$ to construct $U$ and $V$ according to the model complexity; in general, the larger the value of $k$, the smaller the RMSE value between the original matrix $A$ and produced matrix $R$. We call $k$ depth of the decomposed matrix. We denote $R_k=U_k V^\mathsf{T}_k$, and 
we use $S$ instead of $R_k$ to all $k$ here for simplicity. 

We can construct the continuous response matrix with depth of $k$ by using the matrix decomposition method to the observed response matrix; that is $s_{ij}=\sum_{l=1}^k u_{il} v_{jl}$. If we transform $s_{ij}$ to 0/1 discrete value such that $t_{ij}=0, \ (s_{ij} < 0.5), \ t_{ij}=1, \ (s_{ij} \ge 0.5)$, we can construct the discrete response matrix.

\section{Matrix Decomposition Perspective for Accuracy Assessment of Item Response Theory}

\subsection{Measuring the closeness between the two matrices}

We measure the closeness between the two matrices by using two methods; one is the RMSE, and the other is the accuracy.

The RMSE between a matrix $A=(a_{ij})$ and a matrix $B=(b_{ij})$ is defined as below
\begin{eqnarray}
\text{RMSE}=({1 \over \sum_{i=1}^m \sum_{j=1}^n I(i,j)} 
\sum_{i=1}^m \sum_{j=1}^n I(i,j)(a_{ij} - b_{ij})^2 )^{1/2},
\end{eqnarray}
where $a_{ij}$ and $b_{ij}$ are real numbers in $\mathbb{R}$, 
and $I(i,j)=0 \text{\ if element } (i,j) \text{\ is null, }I(i,j)=1 \text{\ if element } (i,j) \text{\ is not null}$.

Since the Frobenius matrix norm and the RMSE
are directly corresponding to each other,
we can expect how close the matrix $A$ is to $B$ via the value of the matrix norm; the closer the matrix $A$ to $B$, the smaller the matrix norm of $A-B$, which corresponds that the closer the matrix $A$ to $B$, the smaller the RMSE of $A-B$. The difference between the Frobenius matrix norm value and the RMSE value is the difference between the root squared error and the root mean squared error.

The accuracy between a matrix $A=(a_{ij})$ and a matrix $B=(b_{ij})$ is defined as below
\begin{eqnarray}
\text{accuracy}={1 \over \sum_{i=1}^m \sum_{j=1}^n I(i,j)} 
\sum_{i=1}^m \sum_{j=1}^n I(i,j)(a_{ij} - b_{ij})^2,
\end{eqnarray}
where $a_{ij}$ and $b_{ij}$ are discrete numbers in $\{0, 1\}$, 
and $I(i,j)=0 \text{\ if element } (i,j) \text{\ is null, }I(i,j)=1 \text{\ if element } (i,j) \text{\ is not null}$.

\subsection{Accuracy assessment} 

Using the maximum likelihood estimates obtained by the item response theory, we can reconstruct the estimated item response matrix, then we can measure how close the reconstructed response matrix to the observed response matrix. The closeness between the two matrices. i.e., the performance is measured by using the RMSE and the accuracy value.

To compare the performance of the IRT with others, we have provided the two kinds of low rank response matrix by approximating the observed response matrix; one is the matrix via the SVD method when the response matrix is a complete matrix, and the other is the matrix via the matrix decomposition method when the response matrix is an incomplete matrix. 
Table \ref{table:Completematrixcase} shows the RMSE and accuracy values between the reconstructed response matrix and the observed response matrix when the observed response matrix is a complete matrix.

We can see, firstly, that the performance of the SVD method and the matrix decomposition method are almost the same both in the RMSE measure and the accuracy measure when the response matrix is a complete matrix. This means that the matrix decomposition method is well catching up the SVD result. 

Secondary, we see that the closeness of the reconstructed matrix obtained from the IRT to the observed matrix is located between the two approximated low rank response matrices obtained from the SVD of $k=1$ and $k=2$ to the observed matrix, where $k$ indicates the first $k$ columns use in the decomposed matrices.

This is amazing in a sense of matrix approximation. 
The reconstructed response matrix using the IRT, whether continuous or discrete, would not exceed the approximated low rank response matrix of $k=2$ using the SVD or the matrix decomposition method.

\begin{table}[htb]
\caption{RMSE and accuracy between the reconstructed continuous response matrix and the observed response matrix when the response matrix is a complete matrix.}
\label{table:Completematrixcase}
\begin{center}
\begin{tabular}{c|c|c|c|c}
\hline
     &    $k$ & SVD & Matrix Decomposition & IRT\\
\hline
RMSE      &    1     &  0.4066 & 0.4067 &\\
                &           &              &             & 0.3915\\
                &    2     &  0.3851 & 0.3854 &\\
                &    3     &  0.3652 & 0.3656 &\\
                &   10   &  0.2563 & 0.2583 &\\
                &   31   &  0.0000 & 0.05570 &\\
\hline
\hline
 accuracy     &     1     &  0.7514 & 0.7527 &\\
                    &            &            &           & 0.7630\\
                    &    2     &  0.7740 & 0.7749 &\\
                    &    3     &  0.7941 & 0.7941 &\\
                    &    10   &  0.9198 & 0.9211 &\\
                    &    31   &  1.0000 & 1.0000 &\\
\hline
\end{tabular}
\end{center}
\end{table}

\begin{figure}[htb]
\begin{center}
\includegraphics[scale=1.2]{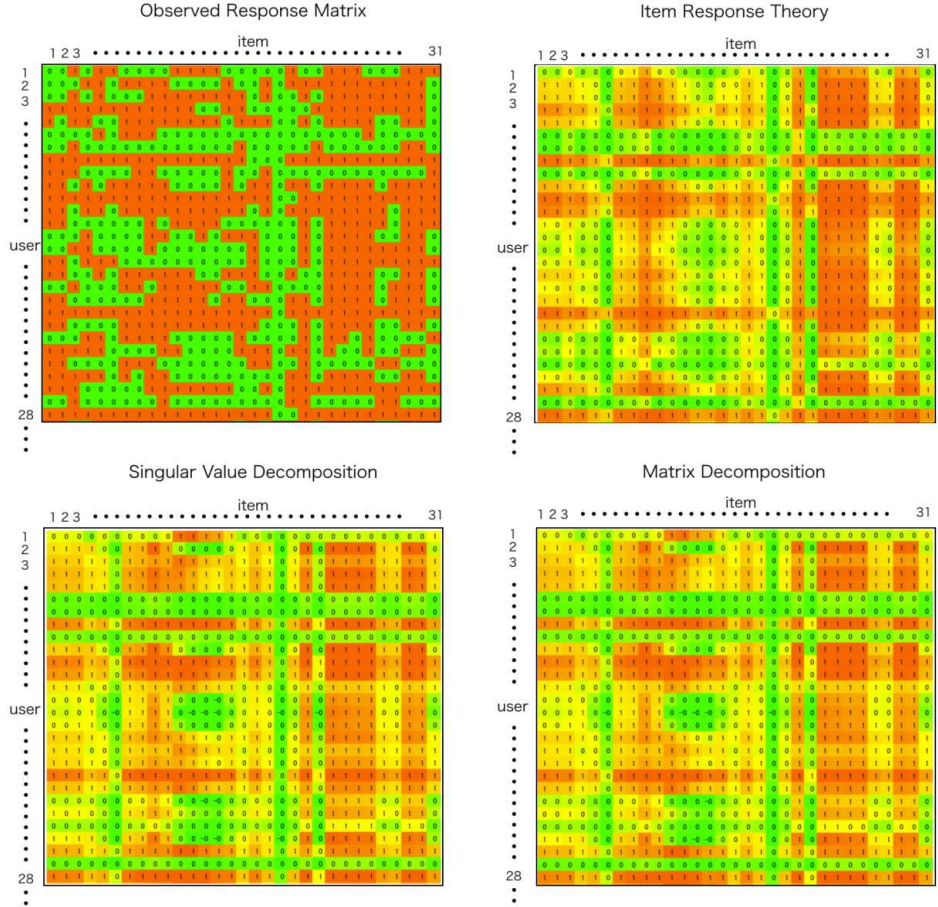}
\end{center}
\caption{Comparison of reconstructed continuous response matrix between IRT and SVD. In SVD and the matrix decomposition, $k=2$.} 
\label{fig:continuousmatrix}
\end{figure}

\begin{figure}[htb]
\begin{center}
\includegraphics[scale=1.2]{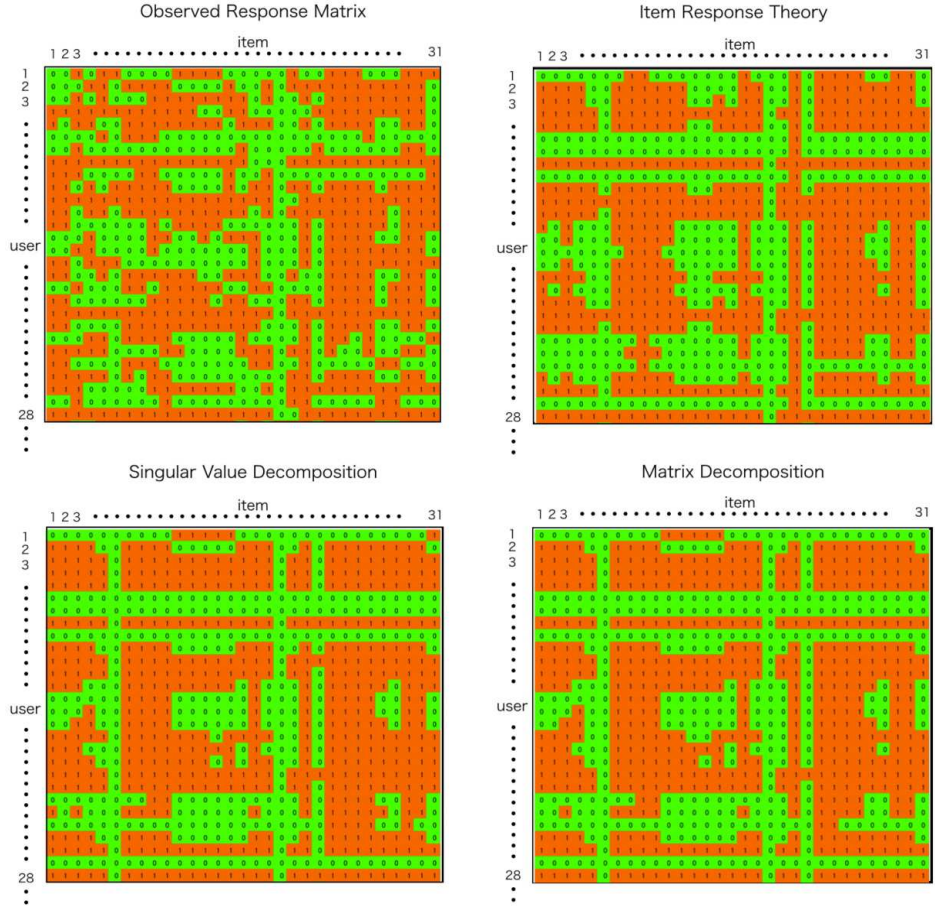}
\end{center}
\caption{Comparison of reconstructed discrete response matrix between IRT and SVD. In SVD and the matrix decomposition, $k=2$.} 
\label{fig:discretematrix}
\end{figure}

Since the reconstructed response matrix is close to the approximated response matrices from the SVD and the matrix decomposition with $k=2$, we have compared the matrices among the originally observed response matrix, reconstructed response matrix estimated by using the IRT, approximated response matrix from the SVD, and approximated response matrix from the matrix decomposition method in Figures \ref{fig:continuousmatrix} and \ref{fig:discretematrix}; Figure \ref{fig:continuousmatrix} compares the continuous case, and Figure \ref{fig:discretematrix} discrete case. It is interesting that the result from the IRT is very similar to those from the SVD and the matrix decomposition in the case of $k=2$.

\section{Incomplete Matrix Case}

As mentioned in the introduction section, we sometimes encounter the case that the observed item response matrix is incomplete. Using the IRT, we can still estimate the maximum likelihood estimators for parameters and reconstruct the response matrix in such a situation by modifying the estimation methodology. However, we cannot construct the response matrix by using the SVD. Instead, the matrix decomposition method, or other methods, e.g., the imputation method with EM algorithm, can be used. Since the constructed response matrices by the SVD and the matrix decomposition are almost the same, we could expect a similar result also in the incomplete matrix case.

\begin{figure}[htb]
\begin{center}
\includegraphics[scale=1.0]{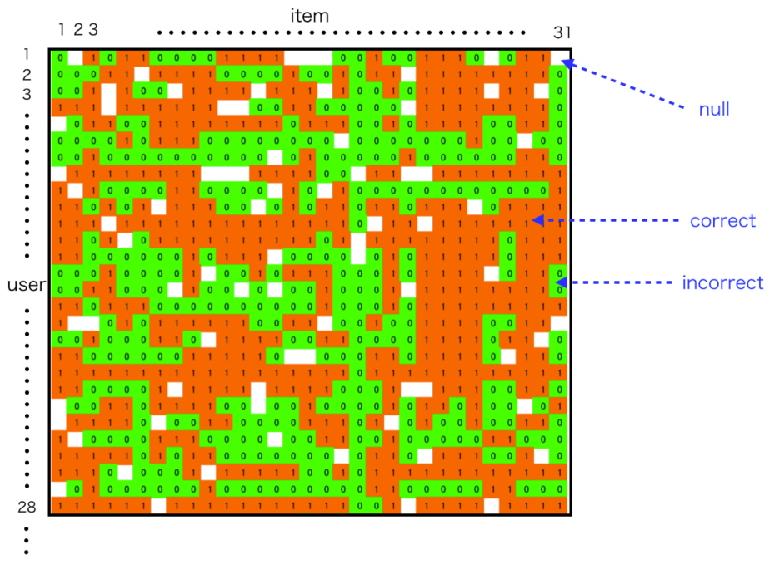}
\end{center}
\caption{Incomplete observed item response matrix} 
\label{fig:incompletematrix} 
\end{figure}

Figure \ref{fig:incompletematrix} shows an artificial incomplete matrix which is made by extracting elements at random from the original matrix. The ratio of the null elements is 10\%.  
Table \ref{table:incompletematrix} shows the RMSE and accuracy values between the reconstructed response matrix and the observed response matrix when the observed response matrix is an incomplete matrix.
As expected, the RMSE and accuracy values between the constructed response matrix and the observed response matrix show similar values provided in Table \ref{table:Completematrixcase}.  
The closeness of the reconstructed matrix obtained from the IRT to the observed matrix is located between the two approximated low rank response matrices obtained from the matrix decomposition method of $k=1$ and $k=2$ to the observed matrix, where $k$ indicates the first $k$ columns use in the decomposed matrices.

\begin{table}[htb]
\caption{RMSE and accuracy between the reconstructed continuous response matrix and the observed response matrix when the response matrix is an incomplete matrix.}
\label{table:incompletematrix}
\begin{center}
\begin{tabular}{c|c|c|c|c}
\hline
  &     $k$ & SVD & Matrix Decomposition & IRT\\
\hline
RMSE     &     1     &  - & 0.4063 &\\
    &             &              &             & 0.3909\\
     &     2     &  - & 0.3834 &\\
     &     3     &  - & 0.3624 &\\
     &     10   &  - & 0.2466 &\\
     &     31   &  - & 0.05309 &\\
\hline
\hline
accuracy     &     1     &  - & 0.7564 &\\
    &             &            &           & 0.7614\\
     &     2     &  - & 0.7823 &\\
     &     3     &  - & 0.8020 &\\
     &     10   &  - & 0.9342 &\\
     &     31   &  - & 1.0000 &\\
\hline
\end{tabular}
\end{center}
\end{table}

\section{Discussions}

\subsection{Matrix rank and the number of free parameters}

The number of unknown parameters to be estimated in the IRT is $m+2n$ where $m > 0$ is the number of users and $n > 0$ is the number of items. In the matrix decomposition scheme, the number of parameters to be estimated is $k(m+n)$. Since $1(m+n) < 1(m+n)+n < 2(m+n)$, The number of free parameters in the IRT, which can also be regarded as the model complexity, is located between the matrix decomposition model of $k=1$ and $k=2$ in a model complexity sense. 

If the function $P$ in the IRT using parameters of $\theta$, $a$, and $b$ were a linear function, it is natural to think that the approximation accuracy of the reconstructed response matrix to the observed response matrix is located between the matrix decomposition model of $k=1$ and $k=2$. However, the function $P$ is nonlinear. Despite of this, the approximated accuracy still lies between the model of $k=1$ and $k=2$. This is surprising. The reason why we surprise is just we merely have not been taking a view from the the matrix decomposition perspective so far. However, it is important and valuable that we can understand the IRT from other viewpoints. This paper does not insist to show the superiority or inferiority of the model performance. The crucial point is the perspective from the matrix decomposition. This is new. However, you might be suspicious to our claim with just one example. We will show other cases in the following.

\subsection{Other example cases}

We have already investigated the difference of the estimates for ability parameters between the mid-term examination case and the end-term examination case before\cite{LTLE2021} using the IRT. Although the ability estimates of the mid-term examination and the end-term examination cases show similar tendencies, it is interesting that the estimates for ability in the mid-term examination are not covered with the 95\% confidence intervals for ability in the end-term examination with probability of 95\%. This is because the questions are different in two term examinations, and some examinees may made progress and some other examinees may not. Thus, the fluctuations of the estimates for ability in the mid-term and end-term results are not the same. We can deal the end-term examination case as a different example case from the mid-term examination.

Table \ref{table:othercases} shows the RMSE between the reconstructed continuous response matrix and the observed response matrix using 12 different examination cases, with various number of examinees and number of questions, including the mid-term and end-term examination cases described above; they are named as case 1 and 2. Since we have seen that the results from the SVD and the matrix decomposition methods are almost the same, we show only the case of the RMSE. As a result, the tendency we have explained above remains the same in all the different example cases. 
We could mention that the RMSE in the IRT model is located between those of the matrix decomposition model of $k=1$ and $k=2$ in a model complexity sense.

\begin{table}[htb]
\caption{RMSE between the reconstructed continuous response matrix and the observed response matrix.}
\label{table:othercases}
\begin{center}
\begin{tabular}{c|c|c|c|c}
\hline
              case  & SVD1 \ \ \ \ \ \  IRT  \ \ \ \ \ \   SVD2  &  $n$ & $m$  & examination  \\
\hline
                1     &  0.4066 $>$ 0.3915  $>$ 0.3851 &  216  &  31 & mid-term \\
                2     &  0.3910 $>$ 0.3790  $>$ 0.3728 &  215  &  36 & end-term \\
                3     &  0.2809 $>$ 0.2688  $>$ 0.2636 &  68  &  42  \\
                4     &  0.3259 $>$ 0.3176  $>$ 0.3062 &  45  &  42  \\
                5     &  0.3974 $>$ 0.3677  $>$ 0.3597 &  171  &  36  \\
                6     &  0.2691 $>$ 0.2509 $>$ 0.2347  &  40  &  19  \\
                7     &  0.3593 $>$ 0.3396 $>$ 0.3313  &  1131  &  22  \\
                8     &  0.3911 $>$ 0.3696 $>$ 0.3687  &  1131  &  55  \\
                9     &  0.2816 $>$ 0.2716 $>$ 0.2432  &  39  &  19  \\
                10   &  0.2473 $>$ 0.2165 $>$ 0.2079  &  20  &  24  \\
                11     &  0.4319 $>$ 0.4162 $>$ 0.3940  &  39  &  50  \\
                12     &  0.4140 $>$ 0.3937 $>$ 0.3863  &  144  &  43  \\
\hline
\end{tabular}
\center{SVD1: $k=1$,  SVD2: $k=2$}
\end{center}
\end{table}

\section{Concluding Remarks}

The item response theory is commonly used as the standard method to evaluate abilities of examinees and difficulties of problems simultaneously in testing. We can obtain the estimates and their confidence intervals for parameters by using the maximum likelihood estimation method when the observed item response matrix is provided. 

Using the maximum likelihood estimates, we can reconstruct the estimated item response matrix. Then we can assess the accuracy of this reconstructed matrix to the observed response matrix from the matrix decomposition perspective. 
That is, this paper focuses on the performance of the reconstructed response matrix. 

To compare the performance of the item response theory with others, we provided two kinds of the low rank response matrix by approximating the observed response matrix; one is the matrix via the singular value decomposition method when the response matrix is a complete matrix, and the other is the matrix via the matrix decomposition method when the response matrix is an incomplete matrix. 

We have, firstly, found that the performance of the singular value decomposition method and the matrix decomposition method are almost the same when the response matrix is complete. Here, the performance is measured by the closeness between the two matrices using the root mean squared errors and the accuracy. Secondary, we have seen that the closeness of the reconstructed matrix obtained from the item response theory to the observed matrix is located between the two approximated low rank response matrices obtained from the matrix decomposition method of $k=1$ and $k=2$ to the observed matrix, where $k$ indicates the first $k$ columns use in the decomposed matrices. 

This is amazing. The reconstructed response matrix using the item response theory, whether continuous or discrete, would not exceed the approximated low rank response matrix of $k=2$ using the singular value decomposition method or the matrix decomposition method, even if the probability function is a nonlinear function of parameters.
This perspective has been supported by many example cases.

\end{document}